\documentstyle[aps,twocolumn,floats]{revtex}
\begin{document}
\draft
\baselineskip 4.0mm
\newcommand{\bec}{\begin{center}}
\newcommand{\ec}{\end{center}}
\newcommand{\be}{\begin{equation}}
\newcommand{\ee}{\end{equation}}
\newcommand{\beqn}{\begin{eqnarray}}
\newcommand{\eeqn}{\end{eqnarray}}
\newcommand{\bet}{\begin{table}}
\newcommand{\ent}{\end{table}}
\newcommand{\bib}{\bibitem}

\title{Perturbation Theory on Top of
Optimized Effective Potential Method}

\author{P. S\"ule} 
\address{Department Natuurkunde,
RUCA, TSM Group\protect\\ Groenenborgerlaan 171., Antwerpen, Belgium\protect\\ 
 E-mail: sule@ruca.ua.ac.be}

\date{\today}

\maketitle

\begin{abstract}

 We present a perturbative approach within the scope of Kohn-Sham density functional theory (DFT). 
The method is based on the exact exchange-only
optimized effective potential method, and correlation is included via
perturbation expansion using Rayleigh-Schr\"odinger (RS) perturbation theory (PT).
The correlation potential is constructed from M$\o$ller-Plesset formulation of RS calculations.
This naturally leads to a new iterative scheme
when finite order perturbation theory is employed. 
The new iterative procedure can be taken as a self-consistent parameter-free DFT PT, and, as such, provides correlation energy
which is explicitly functional of the self-consistent orbitals, eigenvalues
and also of the exchange-correlation potential.
As a demonstration, terms up to the second order in the PT expansion
are considered and the explicit formula for the second order correlation
potential is given as well.

\end{abstract}

\section{Introduction}

 Kohn-Sham (KS) density functional theory (DFT) is a formally exact treatment
of the many-body problem \cite{DFT}. 
Kohn and his co-workers showed that the exact equations of quantum mechanics, taking into account
all the electrons in a molecule (or solid) can be replaced by simpler
equations in which only the density of electrons at each point in space enters
\cite{DFT}.
Remarkable success 
has been achieved in the last decades in finding the "missing link" of KS-DFT, the accurate exchange-correlation
energy functional $E_{xc}[\rho]$. The possibility of further improvements
of model functionals, however, seems to be rather limited within the
scope of the so-called second generation of DFT, in which the
kinetic energy is expressed in terms of orbitals, while the
exchange-correlation energy $E_{xc}$ as an electron density functional.
Furthermore, "state-of-the art" exchange-correlation energy functionals $E_{xc}[\rho]$ and potentials $v_{xc}({\bf r})$,
such as the
generalized gradient approximation (GGA)
\cite{Perdew}
suffer from serious problems since they do not exhibit the correct
$-\frac{1}{r}$ long-range and quadratic behaviour at small $r$ 
\cite{KLI,Sule}.
It became also apparent that GGA is not able to reproduce  
exchange-correlation energies and potentials simultaneously with the
accuracy provided by the energy functionals \cite{Sule}.
All this indicates that although density gradient treatments have had some
success, they are not sufficiantly well developed at present to provide
a final answer to the problem of determining the exchange-correlation
contribution to the ground-state energy functional or to the one-body potential
\cite{March}.

  A new generation of DFT, the optimized effective potential method (OEP)
\cite{Sharp,TS},
opens a new horizon towards fully local and exact 
mean-field theory, where not only the kinetic energy but also the exchange-correlation potential and the exchange-correlation energy are treated at the orbital dependent level of theory. The efficient and highly accurate KLI approximation
to OEP provides an efficient computational tool for the scientific
community. 
OEP and KLI build up correct $-\frac{1}{r}$ assymptotics 
and the highest occupied orbital energy satisfies Koopmans-theorem
and also provides self-interaction free exchange-correlation potential
$v_{xc}({\bf r})$ \cite{KLI}.
 However, the theory is exact at the exchange-only
level and electron correlation can only be added via approximate correlation energy functionals
\cite{Grabo}. It must be emphasized that the correlation energy functionals
neither in their local nor in gradient corrected form are likely to work in combination
with exact exchange. 
This is because basically none of the common correlation functionals have
a long-range component in the corresponding correlation hole (the combined
exchange-correlation hole is typically short-ranged).
The Colle-Salvetti gradient corrected correlation functional \cite{CS}, which
provided excellent results for atoms, performed rather badly
for molecules due to the abovementioned reason \cite{Grabo}.
Approximate correlation functionals, which are derived from the
homogeneous or inhomogeneous electron gas model or obtained non-empirically
from sum-rule conditions, exhibit incorrect long-range tail
(only dynamical correlation is accounted for).
Those functionals also build up
improper local behaviour in atoms and molecules and
their success is mainly due to the cancellation of errors with differrent sign
\cite{Sule,Gritsenko2}.
Therefore, 
the correlation effect can only be calculated in a more complicated
way than in the second generation of DFT 
in those methods which work with exact exchange (e.g. HF and OEP)
\cite{Sulepolym}.

 A few extensions of the OEP method have been made, which allow the
treatment of electron correlation in a limited way.
 In recent years much effort has been devoted to the formulation
of perturbation theory (PT) on KS basis.
 Recently Rayleigh-Schr\"odinger PT has been employed for the DFT correlation energy
using explicit coupling-constant dependence in the model Hamiltonian 
within the KS density functional picture \cite{GL}.
The G\"orling-Levy scheme and OEP provide the exact treatment of the exchange-only
KS problem and also provide an alternative 
to the Hartree-Fock theory.
Casida, on the other hand, employed an approximate perturbative expression
for the ground state to include correlation in OEP \cite{Casida} using
a formalism based on the work of
Sham and Schl\"uter \cite{Sham}.
In a recent review of OEP, Grabo {\em et al.} proposed a scheme which also makes
use of many-body perturbation theory (MBPT) on exchange-only OEP reference state using
Green's functions and the Dyson equation \cite{Graborev}.
Very recently,
Engel {\em et al.}
 worked out the relativistic generalization of OEP and used
M$\o$ller-Plesset based correlation \cite{Engel}.
Aashamar {\em et al.} gave the extension of OEP approach which, permits the treatment
of correlation in a limited way at the multiconfigurational self-consistent
field (MCSCF) level of theory \cite{Talman}. However, this approach met little
practicle relevance being extremely time-consuming 
and also difficult to relate to the single determinent
nature of KS DFT.

  The density functional perturbation theory worked out by X. Gonze {\em et al.}
\cite{Gonze}, which can be taken as a variation-perturbation procedure.
In this as well as in other theories \cite{GL} the self-consistent procedure 
led to the Kohn-Sham wave function being dependent on the perturbation
theory. Thus the perturbation comes into play at an earlier stage than in
conventional quantum chemical perturbation theory (e.g. M$\o$ller-Plesset 
perturbation theory).
Also, Holas and March expressed the exact exchange-correlation potential
in terms of first- and second order density matrices 
using the G\"orling-Levy perturbation theory 
\cite{HolasMarch}.
 All these methods appear somewhat complex in detail and it is
presently difficult to predict which one will prove the most convenient for
practical application and implementation \cite{HolasMarch}.

  The Kohn-Sham formalism for ground states is based on the noninteracting
Schr\"odinger equation
\be
 [\hat{T}+\hat{v}_s] \Phi=E_s^{KS} \Phi,
\label{noninteq}
\ee
the KS equation, where $\hat{v}_s$ is the $N$-electron operator which corresponds
to a local multiplicative potential $v_s({\bf r})$ as a consequence of the 
Hohenberg-Kohn theorem \cite{DFT}.
The one-body KS potential $v_s({\bf r})$ is determined, up to an additive
constant, by the requirement that the ground state of the KS Hamiltonian
operator $\hat{T}+\hat{v}_s$, the KS wave function $\Phi$, yields the same
electron density $\rho({\bf r})$ as the ground state of the corresponding
interacting real system.  
$\Phi$ can be composed of $N$ spin orbitals $u_i$ as a single Slater determinant
for nondegenerate systems.
Therefore,
 the following spinrestricted Kohn-Sham single-particle approach is considered for fermionic systems (we have omitted spin-dependency, but the
extension of the theory is straightforward for that case, throughout
a.u. is used):
\be
[-\frac{1}{2} \nabla^2+ v_s({\bf r})] u_i({\bf r})=
 \epsilon_i u_i({\bf r})
\label{kseq}
\ee
where $u_i({\bf r})$ and $\epsilon_i$ are the single particle orbitals and eigenvalues for a fermionic system and
 $v_s({\bf r})$ is the effective Kohn-Sham single particle potential \cite{DFT}.
\be
v_s({\bf r})=v_{ext}({\bf r})+v_H({\bf r})+v_{xc}({\bf r}),
\ee
where $v_{ext}$ is the potential external to the electronic system that includes
the one created by nuclei, $v_H$ is the Coulomb potential due to the
classical electron-electron repulsion and $v_{xc}$ is a nonclassical term,
the exchange-correlation contribution to the
the KS one-body potential.

 In this article we would like to study a more general class of KS
potentials. 
In a general sense, the exchange-correlation part of the KS potential is not a pure density functional
but rather a complicated functional of eigenvalues and of single particle
orbitals
\cite{Sulepolym},
\beqn
v_s([\rho],{\bf r}) = v_s([\rho,\{u_i\},\{\epsilon_i\}],{\bf r}) && \nonumber \\ 
= v_{ext}([\rho],{\bf r}) + v_H([\rho],{\bf r}) && \nonumber \\ +v_{xc}([\rho,\{u_i\},\{\epsilon_i\}],{\bf r})
\eeqn
In addition to this, in the section V we will show that the exchange-correlation energy $E_{xc}$
is even more complicated energy functional being the explicit functional
of the exchange-correlation potential $v_{xc}({\bf r})$ as well,
\be
E_{xc}=E_{xc}[\{u_i,\epsilon_i\},v_{xc}].
\label{excvxc}
\ee
The same conclusion is derived by others \cite{Graborev}.
As demonstrated by G\"orling and Levy \cite{GL}, it is not necesseraly important
to know how the exchange-correlation potential, based on the
eigensolutions of the KS equations, depends on the density. Those potentials
are always implicitly functionals of the density.

To derive $v_{xc}$ formally, we write $E_{xc}$ rigorously
\be
 E_{xc}[\rho] = \frac{1}{2} \int_0^1 d \lambda \int d {\bf r}' d {\bf r}
\frac{\rho({\bf r}) \rho({\bf r}') [g^{\lambda}([\rho];{\bf r},{\bf r'})-1]}
 {\vert {\bf r}-{\bf r}' \vert}, 
\label{Exc}
\ee
where $g^{\lambda}([\rho];{\bf r},{\bf r}')$ can be interpreted as the pair correlation
function of the fictitious system with interaction strength
parameter (coupling constant) $0 \le \lambda \le 1$ and a ground state 
density which is independent of $\lambda$ \cite{DFT,Sule}.
The exchange-correlation potential $v_{xc}$ is formally defined as the
functional derivative
\be
v_{xc}({\bf r},[\rho])=\frac{\delta E_{xc}[\rho]}{\delta \rho({\bf r})}.
\label{funcder}
\ee
Because of the Hohenberg-Kohn theorem, there exists a one-to-one mapping
between the single particle KS potentials $v_s({\bf r})$ and the densities
$\rho({\bf r})$ which guarantees that the functional derivative given by
Eq.~(\ref{funcder}) is defined \cite{DFT}.
However, the functional derivation can not be obtained explicitly,
since one has to derive the following expression \cite{Sule}:
\beqn
v_{xc}({\bf r},[\rho])=\int_0^1 d\lambda \biggm\{ \int d{\bf r}' \frac{\rho({\bf r}') [g^{\lambda}([\rho];{\bf r},{\bf r}')-1]}
{\vert {\bf r}-{\bf r}' \vert} + && \\ \nonumber
 \frac{1}{2} \int d{\bf r}' d{\bf r}^{''} \frac{\rho({\bf r}') \rho({\bf r}^{''})}
{\vert {\bf r}'-{\bf r}^{''} \vert} \frac{\delta g^{\lambda}([\rho];{\bf r}',{\bf r}^{''})}{\delta \rho({\bf r})}
\biggm\}.
\eeqn
The first term of the potential is the so-called {\em potential} part of the
exchange-correlation energy \cite{Sule}, which is identical with Slater's potential
in the exchange-only (x-only) case, i.e. for $g^{\lambda}$ approximated by $g^{\lambda=0}$, and which is exactly known \cite{Slater} 
as an orbital dependent quantity.
 The {\em response} like second term contains the fundamental functional derivative of the pair correlation function which is, however, an unknown functional. 
The functional
derivative Eq.~(\ref{funcder}) can be calulated only for approximate
functionals analitically where explicit dependence on the density $\rho$ is known and therefore to be applied only for the
so-called second generation of DFT \cite{Graborev}. In the third generation of
DFT one uses $E_x[\{u_i\}]$ rather than $E_x[\rho]$
so that not only the kinetic energy but the 
exchange energy is expressed as orbital dependent energy
functionals.
The correlation energy must be determined, though as a density functional
\cite{Sule,Graborev}.
The central equation in the third generation of DFT is still the KS 
Eq.~(\ref{kseq})
\cite{Sulepolym,Graborev}.
At the best of our knowledge
attempts have not yet been made to express the exchange-correlation energy
as a unique orbital-dependent quantity $E_{xc}[\{u_i\}]$ in OEP.
Furthermore, a general theory to be set up, where both the exchange-correlation
potential and the energy are complicated functionals of the self-consistent
solutions of the eigenvalue problem Eq.~(\ref{kseq}) in the spirit of Eqs. (3)
and (4).

  In this work we give an explicit formulation of a perturbation theory
on top of exchange-only OEP.
We derived an explicit expression
for the correlation potential given by
Krieger {\em et al.} \cite{KLI}.
In section II, we give the short summary of OEP formalism, which we use extensively
in the further sections of this article.
We give in sections III-IV, the potential and response part of $v_{xc}$ (see Eq. (7)) exactly.
We formulate a self-consistent perturbation theory on top of
the exchange-only OEP reference state (section V).
We would also like to compare our OEP-PT scheme with other DFT schemes obtained by 
perturbation theory
\cite{GL,Engel} in order to point out the similarities and differences
between them. 

\vspace{5mm}

\section{Optimized Effective Potential Method}

The starting point of the OEP method is the total energy functional
\beqn
\enspace
E_{tot}^{OEP}[\{u_i\}] = \sum_{i=1}^{occ} \int d{\bf r} u_i^*({\bf r}) \biggm(
-\frac{1}{2} \nabla^2 \biggm) u_i({\bf r}) \nonumber \\
+\int d{\bf r} \rho({\bf r}) v_{ext}({\bf r}) 
+ \frac{1}{2} \int d{\bf r} d{\bf r}' \frac{\rho({\bf r}) \rho({\bf r}')}
{\vert \bf{r}-\bf{r}' \vert} 
+ E_{xc}^{OEP}[\{u_i\}],
\eeqn
where, in contrast with ordinary DFT, the exchange-correlation energy is an explicit
functional of orbitals, and, therefore, only implicit functional of the 
density via Eqs. (2)-(3) \cite{KLI,Graborev}.
$i$ is a collective index for all orbital quantum numbers.
The local single-particle potential appearing in Eqs.~(\ref{kseq})-(3)
must be the optimized one yielding orbitals which minimize $E_{tot}^{OEP}[\{u_i\}]$
so that
\be
 \frac{\delta E_{tot}^{OEP}[\{u_i\}]}{\delta v_s({\bf r})} \biggm|_{v_s=v^{OEP}}
=0.
\label{oep}
\ee
As first pointed out by Perdew and co-workers, Eq.~(\ref{oep}) is equivalent
to the Hohenberg-Kohn variational principle \cite{Graborev,Perdew2}.

 The optimized effective potential method (OEP)
 is given by Talman and Shadwick \cite{TS} for getting the exact exchange potential $v_x({\bf r})$.
 Formally, we make use of chain rule for the functional derivative Eq. (6)
to obtain
\beqn
v_{xc}^{OEP}({\bf r},[\rho])&=&\frac{\delta E_{xc}[\{u_i[\rho]\}]}{\delta \rho({\bf r})}
~~~~~~~~~~~~~~~~~~~ \\ \nonumber
&=& \sum_{i=1}^{occ} \int d{\bf r}' \frac{\delta E_{xc}^{OEP}[\{u_i\}]}{\delta u_i({\bf r^{'}})}
\frac{\delta u_i({\bf r'})}{\delta \rho({\bf r})} + c.c.
\eeqn
where $\{ u_i [\rho] \}$ are the orbitals which are, however, implicitly
functionals of the density.
Applying the functional chain rule again and after some algebra one
can get the following integral equation \cite{KLI,GL,Shaginyan}:
\be
\sum_i u_i({\bf r}) \int d{\bf r}' [v_{xc}^{OEP}({\bf r'})-v_i({\bf r'})] G_i({\bf r},{\bf r'})
u_i^*({\bf r'})+c.c.=0
\label{inteq}
\ee
where 
\be
v_i({\bf r})=\frac{1}{u_i^*({\bf r})} \frac{\delta E_{xc}[\{u_i\}]}
{\delta u_i({\bf r})},
\label{vi}
\ee
and $G_i({\bf r},{\bf r}')$ is the Greens function
\be
 G_i({\bf r},{\bf r}')= \sum_{j \neq i}^{\infty}
\frac{u_j^*({\bf r}) u_j({\bf r}')}{\epsilon_j-\epsilon_i},
\ee
The integral Eq.~(\ref{inteq}) is the fundamental expression for $v_{xc}$ 
in OEP instead of Eq.~(\ref{funcder}), however, there is no known analytic
form of $v_{xc}[\{u_i\}]$. Therefore, only numerical solutions are available
for spherical atoms \cite{KLI,TS,Graborev,Engel2,Talman} and for solids
\cite{Stadele,Kotani,Bylander}.
These numerical solutions of Eq.~(\ref{inteq}) are confined to the exchange-only
OEP and correlation has been taken into account only via approximate local
functionals \cite{Stadele}.
 At the best of our knowledge, the only possibility to account for the correlation energy
exactly is the constrained search formulation of KS theory, when one makes use
of fully correlated density as a reference density obtained from e.g. full CI
calculations \cite{Sule,Gritsenko2}. Although such a calculation provides
exact $v_{xc}$, it represents an enormous computational task which can hardly be
carried out for systems of practical interest.
  G\"orling and Levy have pointed out that Eq.~(\ref{inteq}) is the special case of
a more general equation \cite{GL2}.

 In the work of Krieger and co-workers, the OEP integral equation is analyzed
and a simple approximation is made reducing the complexity of the
original OEP equation significantly, and, at the same time, keeping many of
the essential properties of OEP unchanged \cite{KLI}.
Krieger, Li and Iafrate gave an exact expression by transforming the OEP
integral Eq. (7) into a manageable form. They have got the following, still {\em exact}
expression for $v_{xc}({\bf r})$:
\beqn
v_{xc}^{OEP}({\bf r})=v_{xc}^S({\bf r})+\sum_i^{occ} \frac{\rho_{i}({\bf r})}{\rho({\bf r})}
(\overline{v}_{xci}^{OEP}-\overline{v}_i) && \\ \nonumber
+\frac{1}{2} \sum_i^{occ} \frac{\nabla [p_i({\bf r}) \nabla u_i({\bf r})]}
{\rho({\bf r})},
\label{exactvxc}
\eeqn
where $v_{xc}^S({\bf r})$ is the Slater's potential given as the first term in Eq. (7) in exchange-only case. However, Slater's potential can be generalized
to include the potential part of the correlation energy density as well 
\cite{KLI,Sule}.
The exchange component of $v_{xc}^S$
\be
v_x^S({\bf r}) = - \frac{1}{2\rho({\bf r})} 
\sum_{i,j}^{occ} u_i({\bf r}) 
u_j^*({\bf r}) \int d{\bf r}' \frac{u_i^*({\bf r}') 
u_j({\bf r}')}{\vert {\bf r} - {\bf r}' \vert}.
\label{slater}
\ee

The summation runs over the orbital index for all the occupied orbitals up to the
highest occupied $m$th orbital (Fermi level).
The function $p_i({\bf r})$ is defined by 
\be
p_i({\bf r})=\frac{1}{u_i({\bf r})} \int d{\bf r}' [v_{xc}^{OEP}({\bf r}')-
v_i({\bf r}')] G_i({\bf r},{\bf r}') u_i({\bf r}'),
\ee
with the partial density $\rho_i({\bf r})=u_i^*({\bf r}) u_i({\bf r})$.
In practical applications the last term in Eq. (15) turns out to be quite small
in atomic systems and has a small effect only on the atomic shell boundaries \cite{KLI}. 
The neglect of this term then leads to the KLI-approximation which has the
following form after some algebra \cite{KLI}.
\be
v_{xc}^{KLI}({\bf r})=v_{xc}^S({\bf r})+
\sum_{i=1}^{m-1} \frac{\rho_i({\bf r})}{\rho({\bf r})} \sum_{j=1}^{m-1} ({\bf A}^{-1})_{ij} (\overline{v}_{xcj}^S-\overline{v}
_j).
\label{kli}
\ee
\beqn
\sum_{j=1}^{m-1} ({\bf A}^{-1})_{ij} (\overline{v}_{xcj}^S-\overline{v}
_j) =
\overline{v}_{xci}^{KLI}-\overline{v}
_i, \nonumber
\eeqn
\beqn
{\bf A}_{ji}=\delta_{ji}-{\bf M}_{ji},
\nonumber
\eeqn
\beqn
 {\bf M}_{ji}=\int \frac{\rho_j({\bf r}) \rho_i({\bf r})}
{\rho({\bf r})} d{\bf r},  i,j=1,...,m-1
\nonumber
\eeqn
\beqn
v_{xc}^S({\bf r})=\sum_{i=1}^m \frac{\rho_i({\bf r})}{\rho({\bf r})}
v_{i}({\bf r}),
\nonumber
\eeqn
\beqn
\overline{v}_j^S=\int d{\bf r} \rho_j({\bf r}) v_{xc}^S({\bf r}) 
\nonumber
\eeqn
\beqn
\overline{v}_{xcj}=
\overline{v}_{xj}^{HF}+\int d{\bf r} \rho_j({\bf r})
 v_{cj}({\bf r})
.
\nonumber
\eeqn
\be
\overline{v}_{xj}^{HF} 
 = -\frac{1}{4} \sum_{i=1}^{occ} \int d{\bf r} d{\bf r}' \frac{u_i({\bf r})
u_j^*({\bf r^{'}}) u_i^*({\bf r'}) u_j({\bf r})}
{\vert {\bf r}-{\bf r}' \vert},
\ee
where $m$ is the Fermi level (the highest occupied one-electron energy level)
and note that the $mth$ level is excluded because $\overline{v}_m^{KLI}=
\overline{v}_m^{HF}$ \cite{KLI}.
This might appear a rather crude approximation, but it can be interpreted
as a mean-field approximation since the neglected terms averaged over
ground-state density vanish.
It can be shown that the KLI approximation can be derived alternatively
\cite{Nagy}.


\vspace{6mm}

\section{Perturbation expansion of the correlation energy on top of
exchange-only OEP}

 Let us turn now to the discussion of how to include electron correlation on top
of exact exchange-only KS theory.
Our aim is to solve the exact KS scheme using perturbation
theory on top of a known first order problem (e.g. the x-only OEP).
 Here consider the Hamiltonian for an $N$-electron system
\be
 \hat{\bf {H}}=\hat{\bf {T}}+{\bf \hat{V}_{ee}}+\sum_{i=1}^N v_{ext}({\bf r}_i),
\label{hamil}
\ee
where $\hat{\bf {T}}=\sum_{i=1}^N -\frac{1}{2} \nabla_i^2$,
${\bf \hat{V}_{ee}}=V_{ee}({\bf r})$ stands for the electron repulsion as a local operator and $v_{ext}({\bf r}_i)$
is the external (nuclear) potential of the nuclear frame.
  
 We consider the problem of improving the exchange-only OEP energy
of an $N$-electron system by means of Rayleigh-Schr\"odinger (RS) perturbation theory.
The partition of the Hamiltonian is defined ${\bf H}$ as 
\be
 {\bf H}={\bf H}_0+ {\bf W}
\label{hamil2}
\ee
In particular,
we treat ${\bf W}$ as a perturbation to the {\em exchange-only} KS Hamiltonian ${\bf H}_0$ 
 (${\bf H_0} >> {\bf W}$).
We wish to solve the eigenvalue problem 
\be
\hat{\bf H} \Phi = (\hat{\bf {H}_0}+\hat{\bf {W}}) \Phi =
E \Phi,
\label{eigenveq}
\ee
 which leads to a non-degenerate ground state for a system of $N$-fermions.
$\Phi$ is the ground state wavefunction and $E$ is the expectation value
of the Hamiltonian given by Eq.~(\ref{hamil}) and (\ref{hamil2}).
Eq.~(\ref{eigenveq}) describes the corresponding KS system, a model system of
hypothetical noninteracting electrons with the same ground-state electron density
as the real system \cite{DFT,March,GL}.
However, we only know the solution of the unperturbed problem
\be
 {\bf \hat{H}}_0 \Phi_s =E_0 \Phi_s,
\label{unpert}
\ee
where ${\bf \hat{H}}_0, \Phi_s$ are the unperturbed reference Hamiltonian and single determinent wavefunction.
We assume that Eq.~(\ref{unpert}) provides a complete set of eigenfunctions
$\{u_i\}$ with corresponding eigenvalues $\{\epsilon_i\}$ as well as
the nondegenerate ground energy state $E_0$.
In this article we treat ${\bf H_0}$ as the exact exchange-only KS Hamiltonian,
which is known as the exchange-only OEP Hamiltonian.
Actually, the solutions of exact x-only OEP are equivalent to the solutions
of the exact x-only KS equations \cite{Graborev}.
The central idea is now to determine
the perturbation operator ${\bf \hat{W}}$ as a functional
derivative with respect to the electron density $\rho$ (Eq.~(\ref{funcder})).
\beqn
 {\bf \hat{W}}= W({\bf r})= \frac{\delta (E-\langle \Phi_s \vert {\bf \hat{H}_0} \vert \Phi_s
\rangle)}{\delta \rho} \\ \nonumber
=\hat{V}_{ee}({\bf r})-v_H({\bf r})-v_x({\bf r}).
\eeqn
Therefore, the perturbation operator ${\bf \hat{W}}$ is treated
as a local potential $W({\bf r})$.
 The calculated operator ${\bf \hat{W}}$ can be then substituted back
to the Hamiltonian given by Eq.~(\ref{hamil2}) and the eigenvalue problem
(Eq.~(\ref{eigenveq})) can be solved. The unknown ground state energy $E$ in Eq. (24).
is expressed by RS perturbation theory.
In section V we give the details of this iterative scheme, 
in particular when the reference state is the exchange-only OEP method.

 We treat 
 $v_s([\rho];{\bf r})$ in this article as an explicit functional of the
eigensolutions $\{u_i,\epsilon_i\}$ of the single particle Hamiltonian (to be interpreted as the
KS one). Therefore, the unknown potential $v_{xc}$ is implicit functional of the
ground state density according to the Hohenberg-Kohn theorem \cite{DFT}.
Its uniqueness is also guaranteed by the Hohenberg-Kohn theorem.
 $v_s([\rho];{\bf r})$ in Eq.~(\ref{kseq}) is a one-body operator and Eq.~(\ref{unpert}) reduces
to one-particle equations for KS single particle orbitals $u_i({\bf r})$ as given by
the noninteracting one-particle Schr\"odinger Eq.~(\ref{kseq})
with the exact x-only KS one-body operator $v_s^{xOEP}$.
The KS single determinant, $\Phi_s$, in Eq.~(\ref{unpert}) is then formed
from all occupied one-particle KS orbitals, $\{u_i\}$, i.e., the
$N$ energetically lowest solutions to Eq.~(\ref{unpert}) with 
$N$ being the number of electrons in the system of interest.
Eq.~(\ref{eigenveq}) reduces to exact one-particle KS equations
given by Eq.~(\ref{kseq})
with the exact KS one-body operator $v_s({\bf r})$.
Furthermore, the one-body potentials, which correspond to the x-only reference state ($v_s^{xOEP}$) as
well as to the exact (correlated) case ($v_s^{KS}$) together with the Hamiltonians ${\bf H_0}$ and ${\bf H}$,
are unique functionals of the $v$-representable density.
However, it must be emphasized that the x-only OEP will result in ground
state density different from the exact KS scheme given in Eq.~(\ref{eigenveq}).

 The functional derivation given by Eq. (24) can not be carried out directly
due to the difficulties given in Eqs. (7)-(8). 
The derivative is directly not accessible since $E_{xc}$ is known only in terms
of the one-particle KS states $\{u_i\}$ and the eigenvalues $\{\epsilon_i\}$, and
the explicit functional dependence of the KS eigensolutions on the electron
density is unknown.
 OEP provides an alternative
way of getting the local operator $\hat{\bf W}({\bf r})$.
The solution of Eq.~(\ref{unpert}) is identical with the solution of Eq.~(\ref{inteq}) for exchange-only OEP.
Therefore, we chose the following partition 
of ${\bf H}$ according to Eq.~(\ref{eigenveq}),  
\be
 {\bf H}_0=\sum_i(h_i+\hat{v}_i^{OEP})
\ee
and
\be
  {\bf W}({\bf r})=\hat{V}_{ee}({\bf r}) -\sum_i \hat{v}_i^{OEP}({\bf r}),
\ee
 where $h_i$ is the one-electronic Hamiltonian which contains the kinetic
operator and the operator of the external (nuclear) potential ($\hat{h}_i=\hat{t}_i+
v_{ext}({\bf r_i})$).
$\hat{V}_{ee}$ stands for the electron-electron repulsion as a local operator.
 The exchange component of the perturbed operator is taken into account by
\be
   \hat{v}^{OEP}({\bf r})=\sum_{i=1}^{occ} \hat{v}_i^{OEP}({\bf r}) = \sum_i^{occ} [\hat{j}_i({\bf r})-\hat{k}_i({\bf r})],
\label{exop}
\ee
where $v_H({\bf r})=\sum_i^{occ} \hat{j}_i({\bf r})$ and $\hat{k}_i({\bf r})$ are the corresponding {\em local} Coulomb and exchange
operators. 
$\hat{k}_i({\bf r})$ is given according to Eq. (15),
where $v_x^{OEP}({\bf r})=\sum_{i=1}^{occ} \hat{k}_i({\bf r})$.
The partition of the Hamiltonian given above can be taken as an operator for
M$\o$ller-Plesset (MP) perturbation expansion \cite{MP,Szabo} employed on orbitals
and obtained by the solution of x-only OEP equations. 
 
 The exchange-only OEP total energy $E_{tot}^{xOEP}$ can be considered as the sum of zeroth and first-order
energies,
\beqn
 E_{tot}^{xOEP}=E^{(0)}+E^{(1)}= \sum_i^{occ} \epsilon_i - \langle \Phi_s \vert {\bf \hat{W}} \vert \Phi_s \rangle && \\ \nonumber 
= \sum_i^{occ} \epsilon_i - E_x^{OEP}[\{u_i\}],
\eeqn
where
\be
 E_x^{OEP}[\{u_i\}]=-\frac{1}{4} \sum_{ij}^{occ} \int d{\bf r} d{\bf r}' \frac{u_i({\bf r})
u_j^*({\bf r}) u_i^*({\bf r}') u_j({\bf r}')}
{\vert {\bf r}-{\bf r}' \vert}.
\label{exoep}
\ee
According to Eq. (9)
the exact (correlation corrected) total energy $E_{tot}^{(n),OEP}$ can be given at 
the $n$th order level of perturbation theory
\be
 E_{tot}^{(n),OEP}[\{ u_i\}]=E_{tot}^{xOEP}+E_c^{(n \ge 2)}.
\ee
The first correction to the exchange-only energy occurs in the second order of
perturbation theory.

 The correlation energy is expanded as the sum of
correlation contributions at any order, so that
\be
E_c^{(n)}[\{u_i,\epsilon_i\}] =\sum_{p=2}^n \varepsilon^p[\{u_i,\epsilon_i\}].
\label{ecexpan}
\ee

 Next step is to express the unknown correlation potential in order to
solve the correlated KS problem given by Eq.~(\ref{eigenveq}).
The correlation potential is also determined according to the perturbation
theory at $n$th order level of theory keeping in mind that correlation and
exchange effects are decomposed,
\be
v_{xc}({\bf r})=v_x({\bf r})+v_c^{(n)}({\bf r}).
\ee
For $v_c$
we employ the expression, obtained from the exact transformation of the OEP
integral Eq.~({\ref{inteq}) and given by Eq. (15).
For sake of simplicity, however, we use the remarkably accurate KLI approximation,
and neglect the rather difficult last term in Eq. (15). 
According to Eq.~(\ref{kli}) $v_c^{(n)}$
can then be written 
\be
v_c^{(n)}({\bf r})=v_c^{S(n)}({\bf r})+\sum_{ij}^{m-1} \frac{\rho_i({\bf r})}{\rho({\bf r})}
{\bf A}_{ij}^{-1} (\overline{v}_{cj}^{S(n)}-\overline{v}_{cj}^{(n)}).
\label{vc2}
\ee
{\em We would like to note here that none of the existing DFT-PT schemes \cite{GL,Engel,Gonze} used 
directly this OEP expression for the correlation potential.}
Instead they derived the correlation potential in an alternative way.
G\"orling and Levy employed functional derivation over the second order
correlation energy with respect to the Kohn-Sham potential and to the
eigenvalues. Engel {\em et al.} also applied the GL PT 
 over the relativistic OEP leading to functional derivations with respect to
the KS orbitals and to the KS eigenvalues. 
Both theories will lead to different expressions.
We use, however,
functional derivation with respect to only the orbitals according to Eq. (13). 
In the next few steps and in the next section we will show that this difference
will lead naturally to different correlation potential then those given
by the authors mentioned above.
This formulation of the correlation potential has {\em Slater} and {\em response} component \cite{Sule,KLI},
\be
v_c^{(n)}({\bf r})=v_c^{S,(n)}({\bf r})+v_c^{resp,(n)}({\bf r}), 
\ee
and can be extracted as follows,
\be
v_c^{S(n)}({\bf r})=\sum_{j=1}^{occ} \frac{\rho_j({\bf r})}{\rho({\bf r})}
v_{cj}^{(n)}({\bf r}),
\label{vcS}
\ee
\be
\overline{v}_{cj}^{S(n)}=\int d{\bf r} \rho_j({\bf r}) v_c^{S(n)}({\bf r}),
\ee
\be
\overline{v}_{cj}^{(n)}=\int dr \rho_j({\bf r}) v_{cj}^{(n)}({\bf r}),
\ee
The $n$th-order orbital dependent potential $v_{cj}^{(n)}$ can be given using the 
M$\o$ller-Plesset correlation energy $E_c^{(n)}$ given in Eq.~(\ref{ecexpan})
\be
 v_{cj}^{(n)}({\bf r})=\frac{1}{u_j^*} \frac{\delta E_c^{(n)}}{\delta u_j}.
\label{vcfuncder}
\ee
In the next section we will show that the functional derivation with respect
to the orbitals can be given analitically and finally one can get closed
form for the correlation potential at $n$th-order level of PT.

 According to Eqs. (22-26)
the following self-consistent Kohn-Sham procedure can then be
constructed, which is correlated at the $n$th order level of RS
PT,
\beqn
[-\frac{1}{2} \nabla^2+ v_{ext}({\bf r})+v_H({\bf r})+ v_x({\bf r})+
v_c^{(n)}({\bf r})] u_i({\bf r}) && \\ \nonumber =\epsilon_i u_i({\bf r}).
\eeqn
We will discuss the various properties of this self-consistent KS PT scheme
in section V.

\section{The correlation energy and potential at second order}

Because of the increasing complexity of perturbation expansion,
 we give here only the MP perturbation energy at 
second order. For a general case one can give (in principle) those quantities at
higher order as well.
Note that one has to consider not only {\em double} excitations but also {\em single}
configurations. In case of HF theory $E_c^{(2),single}=0$ by Brillouin theorem, but in case of
KS self-consistent orbitals we still have to calculate it \cite{Szabo,Warken,FBartha}.
This term does not vanish completely but must be rather small for KS orbitals as well.
For a closed-shell system, the second-order energy can be written in terms
of sums over spatial orbitals as
\beqn
E_c^{(2)}[\{u_i\}]= 
 E_c^{2,single}+E_c^{2,D} && \\ \nonumber
=\sum_{r=1}^2 \frac{\vert \langle \Phi_s \vert
\hat{V}_{ee}({\bf r})-\hat{v}^{OEP}({\bf r}) \vert \Phi_{s,r} \rangle \vert^2}
{E_s-E_{s,r}} && \\ \nonumber
=\sum_i^{occ} \sum_k^{vir} \frac{\vert{\bf W}_{ik}\vert^2}{\epsilon_i-\epsilon_k}
+\sum_{ij}^{occ} \sum_{kl}^{vir} \frac{\vert{\bf W}_{ij}^{kl}\vert^2}
{\epsilon_i+\epsilon_j-\epsilon_k-\epsilon_l}. \\ \nonumber
\eeqn
\beqn
{\bf W}_{ik}=\langle \Phi_s \vert \hat{\bf W} \vert \Phi_i^k \rangle 
=-\sum_j \langle ij \vert ik \rangle-\overline{v}_{x,ik}^{OEP}, 
\nonumber
\eeqn
\beqn
\overline{v}_{x,ik}^{OEP}=\langle i \vert v_x^{OEP}({\bf r}) \vert k \rangle,
\nonumber
\eeqn
\beqn
{\bf W}_{ij}^{kl}=\langle \Phi_s \vert \hat{\bf V}_{ee} \vert \Phi_{ij}^{kl} \rangle=
\vert \langle ij \vert \vert kl \rangle \vert^2, \\ \nonumber
\eeqn
where $\vert \langle ij \vert \vert kl \rangle \vert^2=\langle ij \vert kl \rangle (2 \langle kl \vert ij \rangle -
\langle kl \vert ji \rangle)$,
\be
\langle ij \vert kl \rangle=\int d{\bf r} d{\bf r}'
\frac{u_i({\bf r}) u_j({\bf r}') u_k({\bf r}) u_l({\bf r}')}
{\vert {\bf r}-{\bf r}' \vert}.
\label{ijkl}
\ee
$\Phi_s$ is the $N$-electron Kohn-Sham ground-state single determinent wave function. 
The $r$th excited state to the Kohn-Sham equation is given by $\Phi_{s,r}$.
The expectation values are as follows, $E_s=\langle \Phi_s \vert \hat{H_0} \vert
\Phi_s \rangle$ and $E_{s,r}=\langle \Phi_{s,r} \vert \hat{H} \vert \Phi_{s,r} \rangle$.
Note that since the quantity being summed in Eq.~(\ref{ijkl}) is symmetric
in $i,j,k$ and $l$ it vanishes when $i=j$ or $k=l$.

 The next question to be addressed is the derivation of the
correlation potential $v_c$ according to Eqs. (15) and (18) bearing
in mind the decomposition $v_{xc}=v_x+v_c$.
 The correlation potential at a given order of the perturbation series $v_c^{(n)}({\bf r})$ can be given formally 
by means of a 
functional derivative of the correlation energy $E_c^{(n)}$ with respect to the
orbitals $u_i({\bf r})$ according to Eqs.~(\ref{vi}) and (38)
via the orbital dependent quantity
\beqn
v_{ci}^{(2)}({\bf r})=\frac{1}{u_i^*({\bf r})} \frac{\delta E_c^{(2)}[\{ u_i \}]}
{\delta u_i({\bf r})} \\ \nonumber
=\frac{1}{u_i^*({\bf r})} \biggm\{ \frac{\delta E_c^{(2),single}}{\delta u_i({\bf r})}+
\frac{\delta E_c^{(2),D}}{\delta u_i({\bf r})} \biggm\}.
\eeqn
According to Eqs. (38) and (41) the quantity $v_{ci}^{(n)}$ at second order
\beqn
 v_{ci}^{(2)}({\bf r})= v_{ci}^{(2),single}+v_{ci}^{(2),D}= &&  \\ \nonumber -\frac{2}{u_i^*}
\biggm\{\sum_k^{vir} \frac{\sum_j^{occ} ([j \vert ik]+[ij \vert k])
+[v_x^{OEP} k]+[ik \frac{\delta v_x^{OEP}}{\delta u_i}]}
{\epsilon_i-\epsilon_k} && \\ \nonumber
 +\sum_j^{occ} \sum_{k > l}^{vir} \frac{  
[j \vert kl ]^2}{\epsilon_i + \epsilon_j -\epsilon_k - \epsilon_l}
\biggm\},
\eeqn
since
\beqn
[j \vert kl ]^2=
\frac{\delta}{\delta u_i({\bf r})} \vert \langle ij \vert \vert kl \rangle \vert^2 = [j \vert kl ] 
 (2\langle kl \vert ij \rangle -\langle kl \vert ji \rangle) && \\ \nonumber + [kl \vert j] \langle ij \vert kl \rangle 
\nonumber
\eeqn
\beqn
 [j \vert kl] = u_k({\bf r}) \int d{\bf r}' \frac{u_j({\bf r'})
 u_l({\bf r'})}{\vert {\bf r}-{\bf r'} \vert}, \nonumber
\eeqn
The Slater component of the second order correlation potential given by Eq.~(\ref{vcS}) will be
precisely
\be
v_c^{S,(2)}({\bf r})=v_c^{S,single,(2)}({\bf r})+v_c^{S,D,(2)}.
\ee
Using the decomposition $v_{xc}=v_{xc}^S+v_{xc}^{resp}$,
the potential part of $v_{xc}$ or equivalently the generalized Slater component 
of Eq.~(\ref{kli}) at second order of PT can be finally written (note that only {\em double}
excitations are included, $v_c^{single,(2)}$ is given in the Appendix),
\beqn
v_{xc}^{S,D,(2)}([\rho,\{u_i,\epsilon_i\}];{\bf r})=v_x^S({\bf r})+v_c^{S,D,(2)}({\bf r}) = && \\ \nonumber
 \frac{1}{\rho} \biggm( \sum_{ijkl}^{occ} [ij \vert kl] +
\sum_{ij}^{occ} \sum_{kl}^{vir} 
\frac{[ij \vert kl]^2}{\epsilon_i+\epsilon_j-\epsilon_k
-\epsilon_l} \biggm),
\eeqn
and the generalized response part
\beqn
 v_{xc}^{resp,D,(2)}([\rho,\{u_i,\epsilon_i\}];{\bf r})=
 v_x^{resp}({\bf r})+v_c^{resp,(2)}({\bf r})=&& \nonumber \\
 \sum_{ij}^{m-1} \frac{\rho_i}{\rho}
{\bf A}_{ij}^{-1} \biggm\{ \sum_{ikl}^{occ} \int d{\bf r} 
 \frac{\rho_j}{\rho} \sum_j^{occ} [ij \vert kl]-\frac{1}{4} \langle ij \vert kl \rangle)+ && \nonumber \\
\sum_i^{occ} \sum_{kl}^{vir} 
\biggm( \int d{\bf r} \frac{\rho_j}{\rho} \sum_j^{occ} \frac{[ij \vert kl]^2}{\epsilon_i+\epsilon_j-\epsilon_k
-\epsilon_l} && \nonumber \\  
-\frac{\vert \langle ij \vert \vert kl \rangle \vert^2}{\epsilon_i+\epsilon_j-\epsilon_k
-\epsilon_l} \biggm) \biggm\}. 
\eeqn
The expression is given only up to the second order, and the extension of the
formula to include higher order terms is straightforward, although order-by-order with increasing complexity \cite{Szabo}.


\section{The self-consistent perturbation scheme}

 Having discussed the partition scheme and the formulation of $n$th-order xc-potential,
 we would like to study the procedure given by Eq. (39). This scheme allows one to improve  
the eigenfunctions and eigenvalues of the unperturbed problem systematically.
The $n$th-order correlation potential
$v_c^{(n)}({\bf r})$ (Eq.~(\ref{vc2})), which is calculated on top of the
first-order level of perturbation theory, is 
substituted back to the exchange-only KS equation 
in order to achieve self-consistency in the field
of $v_c^{(n)}$. 
The introduction of $v_c^{(n)}$ via Eq.~(\ref{vc2}) together with Eq. (39) opens a new variational freedom
in the original KS Eq.~(\ref{kseq}), which leads to a new iterative scheme.
A new $v_c^{(n)}$ can then also be constructed on top of the variationally solved 
correlated KS-problem, however, this would lead to a more complicated iterative
scheme. 

 The iterative scheme given by Eq. (39) is working in the following way: 

 (i) one has to solve the first order level of theory
which is exactly the exchange-only OEP reference state in this particular case
providing the set of orbitals and eigenvalues, $\{u_i\}$ and $\{ \epsilon_i\}$.

 (ii) the next step is the construction of the $n$th-order correlation energy (Eq.~(\ref{ecexpan})) and the exchange-correlation potential
$v_{xc}^{(n)}$ on top of the exchange-only reference state (Eqs.~(\ref{kli}) and (\ref{vc2})).

 (iii) substituting back the   
potential $v_c^{(n)}$ obtained in step (ii) to the KS equation a new correlated KS problem can be solved.
In this way a new set of eigensolution $\{u_i,\epsilon_i\}$
can be obtained.
One can stop at this point, when the $n$th-order correlated KS problem is solved, and
the ground state energy can then be calculated by Eq. (9).

 (iv) Although present paper is addressed to the formulation of a standard perturbation
theory on top of x-only OEP (steps i-iii), we indicate that
further iterative steps can also be considered in practice when $n \neq \infty$. 
The orbitals $\{u_i\}$ and the potential $v_x$ obtained in step (iii) differ from those
obtained in step (i). Therefore, a new $E_c^{(n)}$ and $v_c^{(n)}$ can be calculated which,
however, leads to a KS problem again. This procedure 
continues until convergency is reached, provided, of course, that such a method
turns out to be convergent.
These additional iterative steps with the "updated" $v_c^{(n)}$ result, however,in the change of
the unperturbed problem. This would lead to the change of the partition scheme, the formula given
by Eq. (40) for $E_c^{(n)}$ e.g. at second order of PT. 
Therefore, in Eq. (40) $v^{OEP}({\bf r})$ contains not only $v_x$ 
but also $v_c^{(2)}({\bf r})$ so that,
\be
\hat{W}^{(2)}({\bf r})=\hat{V}_{ee}({\bf r})-\hat{v}^{OEP}({\bf r})+v_c^{(2)}({\bf r}).
\ee

 Here we restrict ourself to the discussion of iterative procedure with only
unchanged partition scheme of the Hamiltonian.
 The more comprehensive discussion of such a RS scheme with improved partition of
the Hamiltonian can be the subject of further studies.

 The solutions of the correlated KS equation obtained in step (iii) can only be taken as
final solutions of Eq. (39) when infinite order PT is employed for
calculating $v_c^{(n)}$ ($n=\infty$).
This is because the RS theory is mathematicaly equivalent representation of
the full configuration interaction (full CI) theory \cite{Szabo}
at infinite order expansion.
In practice, only truncated perturbation expansions can only be considered,
and, as such, fast convergence in the PT expansion is of great
importance.
Therefore, 
the proper choice of the reference space is also important  
to achieve rapid convergence \cite{Chaudhuri}.
Note that in this scheme the reference state is subsequently optimized
when steps (i)-(iv) are used.

 Like in standard Hartree-Fock or KS procedures, the single particle
Eq. (39) have to be solved iteratively until self-consistency is reached.
 According to Eq.~(\ref{vc2})
the scheme can be improved systematically by considering higher order $v_c^{(n)}$.
One of the main differences between the original form of the RS scheme and 
which is
given in Eq. (39) is that in the OEP PT scheme the wave-function and the eigenvalues are improved
via the solution of the $n$th-order KS problem. This theory never goes beyond
the single-determinant picture in the present form.
Exctited configurations are considered, however, when calculating $E_c^{(n)}$
and $v_c^{(n)}$.
This scheme alows one to drive a system progressively from a noninteracting
reference state toward a fully correlated system.
When step (iv) is switched on,
the orbitals $\{u_i\}$ and eigenvalues $\{\epsilon_i\}$ in the
RS expansion of $E_c^{(n)}$ (Eq.~(\ref{ecexpan})) are allowed to be the
updated ones in the RS perturbation expansion. 
The scheme can therefore be considered as a {\em self-consistent} RS perturbation problem (SC-RS OEP). An alternative approach to carry out PT calculation self-consistently is the variation-perturbation approach 
\cite{Gonze,Chaudhuri}.
Within the exact SC-RS OEP scheme self-consistency can be reached 
once the exact $v_{xc}$ is evaluated.
We hope that this iterative scheme might have the "charming feature"
of including the important part of electron correlation already at 
the second order of the correlation energy.
Normally, the eigenvalue problem up to the first order level of theory is
solved variationally. The iterative scheme (39) permits getting
the best optimized first order reference state in RS theory.
The optimization of the reference state is accomplished by the reiteration
of the updated xc-potential and by the consequent solutions of the
"updated" KS problems until convergency is reached in step (iv).
The self-consistent scheme given by Eq. (39) is similar to the so-called
density functional perturbation theory (DFPT) \cite{Gonze} in that respect
that succesive orders of perturbation are obtained iteratively in both cases.
However, DFPT uses energy derivatives with respect to the ordering parameters as a variation-perturbation treatment.

 G\"orling and Levy (GL) \cite{GL,GL2} have given more general functional for $E_{xc}$
and for $v_{xc}$ on the basis of PT. They found on KS basis that both the xc-energy and the
xc-potential must be complicated functionals of both orbitals and eigenvalues
and also of a linear response function type inverse operator $\hat{G}^{-1}$.
However, the quantity $\hat{G}^{-1}$ is analytically unknown.
The GL perturbation theory can provide an exact formal Kohn-Sham
scheme
only in basis set representation.
Both the GL and the OEP PT theories lead to the exact formal representation
of the KS equation, however, the construction of the $n$th-order 
$v_{xc}$ is different.
In the GL PT a coupling constant $\alpha$ dependent scheme links
the noninteracting $N$-electron system with the interacting real system where the electron density
remains independent of $\alpha$ \cite{GL2}. The correlation
potential and the energy are expressed in a Talyor series with respect to the
$\alpha$.
The OEP PT scheme is a coupling-constant free formalism and the
electron density does not remain constant during the perturbation treatment. The exchange-only OEP will result in 
different density from Eq. (39), although the difference may be rather small.
In OEP PT a particular form of $v_{xc}$ is used given by Eq. (15), while
in GL PT a more general, but unknown form of $v_{xc}$ is used.
The main advantage of the present scheme is that the functional derivative
Eq.~(\ref{vcfuncder}) is directly accesible and, therefore, the $n$th-order potential
$v_c^{(n)}$ can be given analitically.
Both theories deliver the exact exchange-correlation energy and potential
order by order.
Holas and March give the leading term of GL PT correlation potential in
terms of first and second order density matrixes (2DM). The 2DM is also expressed
by PT. They also derived an exchange-potential which is free of the
energy denominators $\epsilon_i-\epsilon_k$ while the corresponding
expression in GL theory does contain it \cite{HolasMarch}.
  The extension of relativistic exchange-only OEP with perturbation theory has first been suggested 
by Engel {\em et al.} very recently \cite{Engel}.
 Although their scheme is closely related to our OEP-PT approach, the construction
of the correlation potential seems to be somewhat different.
Not taking into account its relativistic features functional derivatives
appear in their approach with respect to the eigenvalues which are not included in the present
scheme. We use Eq. (33) for the correlation potential together
with Eq. (38) which will lead to somewhat differnet formula than that given
by Engel {\em et al.}. Although they give a very accurate and detailed
formulation of the theory we belive that our approach is still relevant
for applications in the next future.

 In our SC OEP PT scheme the exchange-correlation potential can be given
according to OEP method. The most convenient way is to use the
KLI approximation \cite{KLI} for the xc-potential according to Eq.~(\ref{kli}).
The exchange-only KLI approximation given by Eq.~(\ref{kli}) performs remarkably well \cite{KLI,Graborev}
providing nice agreement with the Hartree-Fock total energies.
In the GL scheme one has a relatively complicated formula at the
first order level of PT, e.g. at the exchange-only level using orbital, eigenvalue and $\hat{G}^{-1}$ dependent exchange-potential. The SC OEP-PT method
uses the much simpler exchange potential according to Eqs.~(15)
and (\ref{exoep}) which is only orbital-functional. 
However, the OEP PT scheme given in Eq. (39) leads to completely exact theory
when the last term in Eq. (15), which is neglected in the KLI approximation,
is calculated as well. Although the integrodifferential equation Eq.~(\ref{inteq})
can be solved only numerically, however, its exact transformation given by
Eq. (15) can be computed exactly for arbitrary fermionic system. 
The neglected term in Eq. (15), when considered, can increase 
the computational difficulties significantly.

 $E_{xc}^{(n)}$ is also  
the functional of the optimized exchange-correlation potential $v_{xc}^{(n)}$ via
the iterative scheme given by Eq. (39),
\be
 E_{xc}^{(n)}=E_{xc}^{(n)}[\{\Phi_s\},\{\epsilon_i\},v_{xc}^{(n)}].
\label{Excvxc}
\ee
This is because in steps (iii) and (iv) in Eq. (39) the correlated
KS problem is  solved using the "updated" $v_c^{(n)}$. 
This equation reflects the complexity of the eigenvalue problem given by
Eq. (38).
Also, Eq.~(\ref{Excvxc}) demonstrates the difference of the SC-RS OEP scheme
from the original formulation of the RS perturbation theory. The expanded
$n$th order correlation energy is explicit functional not only of the self-consistent orbitals and the eigenvalues, but also of the self-consistent
exchange-correlation potential. 
Similar results are reported on Sham-Schl\"uter basis \cite{Sham,Graborev}
and by Engel {\em et al.} \cite{Engel}.
In that and other studies the variational energy expression has been treated as
the functional of the fully dressed Green function $G$ and of the reference
function $G_0$, and they are associated with the Dyson equation \cite{Casida}.

\section{Conclusions}

In this paper we have introduced a first-principle, parameter-free
perturbational density functional scheme in which all exchange-correlation effects are
consistently represented in terms of the eigensolutions of the Kohn-Sham equations.
A self-consistent formulation of perturbation theory is
developed on top of the exchange-only optimized effective potential
(OEP) method.
This generalization of OEP opens
a new variational freedom which leads to a new iterative procedure. The total energy is also the functional
of the $n$th-order exchange-correlation potential. 
In this scheme, the correlation energy and the correlation potential
is expressed via perturbation series while the orbitals and the
 eigenvalues are variationally optimized via the Kohn-Sham equation.
We give exactly the potential and the response
parts of the exchange-correlation potential which are explicitly functionals
of the eigensolutions of the Kohn-Sham eigenvalue problem.
 The correlation potential is given directly as a functional derivative
over the M$\o$ller-Plesset correlation energy expression with respect to the
Kohn-Sham orbitals. 

  We also discuss the difference and similarities between the present and other perturbation
theories which are also based on Kohn-Sham orbitals.

 Further investigations will be fruitful starting from the formalism
introduced concerning the selection of the most appropriate reference
state, the Hamiltonian partition scheme and the convergence of the RS PT based on OEP.
We would like to emphasize the utility of Eq. (39) as a way of improving
Kohn-Sham eigenvalues for calculations of band gaps, ionization potentials
and excitation energies. The self-consistent OEP PT  scheme introduced here can be useful in a wide range
of areas in quantum chemistry and in solid state physics. Whether the present approach
can be really competitive among known high accuracy but costly approaches, such as 
configuration interaction (CI), 
will be hopefully the subject of further tests in the
future.

\vspace{1cm}


 {\small
{\bf Acknowledgments}\\

\vspace{4mm}

 We gratefully acknowledge the useful discussions with \'A. Nagy, F. Bartha
and Professor N. H. March. Also thanks to Dr. Vic Van Doren for his constant support.
This research is supported by the Flamish National Science Foundation.
}

\vspace{1cm}

\begin{center}
{\bf APPENDIX}\\
\end{center}

In order to get the second-order correlation potential one has to
derive the single excitation components which occur when we are dealing
with a "one-body" single particle Hamiltonian.
In Hartree-Fock theory single particle excitations do not play a role
in the second order PT correlation energy formula because
$\langle \Phi_s \vert \hat{W} \vert \Phi_s^{ik} \rangle=0$ where
$\Phi_s^{ik}$ is the single excited determinent wave function.
However, its magnitude
and contribution to the second-order correlation energy
in PT-DFT must be rather small as well. 
Further arguments in this respect can be given only computationally.

  Let us again decompose $v_c^{(2)}$
 into contributions from {\em single} and 
{\em double} excitations, 
\be
v_c^{(2)}({\bf r})=v_c^{single,(2)}({\bf r})+v_c^{D,(2)}({\bf r}).
\ee
$v_c^{D,(2)}$ has been given already in section IV in Eqs. (47) and (48).
Together with Eqs. (33)-(38) $v_c^{single,(2)}$ can be partitioned to Slater
and response parts
\be
v_c^{single,(2)}=v_c^{S,single,(2)}+v_c^{resp,single,(2)},
\ee
\be
 v_{ci}^{(2),single}=\frac{1}{u_i^*} \frac{\delta E_c^{(2),single}}{\delta u_i},
\ee
where
\be
 \frac{\delta E_c^{(2),single}}{\delta u_i}=
-2 \sum_k^{vir} \frac{\sum_j^{occ}([j \vert ik]+[ij \vert k])
+\frac{\delta \overline{v}_{x,ik}^{OEP}}{\delta u_i}}
{\epsilon_i-\epsilon_k}.
\ee
\be
\overline{v}_{x,ik}^{OEP}=\langle i\vert v_x^{OEP}({\bf r}) \vert
k \rangle,
\ee
\be
\frac{\delta \overline{v}_{x,ik}^{OEP}}{\delta u_i}=
\frac{\delta \overline{v}_{x,ik}^S}{\delta u_i}+\frac{\delta \overline{v}_{x,ik}^{resp}}{\delta u_i}
,
\ee
\be
\frac{\delta \overline{v}_{x,ik}^S}{\delta u_i}=
 v_x^S({\bf r}) u_k({\bf r})+ u_i({\bf r}) u_k({\bf r})
\frac{\delta v_x^S}{\delta u_i},
\ee
\be
\frac{\delta v_x^S}{\delta u_i}=
\frac{1}{\rho} \biggm(
\frac{u_i^* e_x^{HF}}{\rho} 
-\frac{\delta e_x^{HF}}{\delta u_i}\biggm)
\ee
where formally the HF exchange energy density is the one given in Eq.~(\ref{exoep})
e.g.,
\be
E_x^{OEP}=\int d{\bf r} e_x^{HF}[\{u_i^{OEP}({\bf r})\}],
\ee
\be
e_x^{HF}=v_x^S({\bf r}) \rho({\bf r}).
\ee
\be
\frac{\delta \overline{v}_{x,ik}^{resp}}{\delta u_i}=
 v_x^{resp}({\bf r}) u_k({\bf r})+ u_i({\bf r}) u_k({\bf r})
\frac{\delta v_x^{resp}}{\delta u_i},
\ee
\beqn
\frac{\delta v_x^{resp}}{\delta u_i}= \frac{\rho_i}{\rho}
\sum_{j=1}^{m-1} \biggm[ \frac{\delta {\bf A}_{ij}^{-1}}{\delta u_i}
(\overline{v}_{xj}^S-\overline{v}_{xj}^{HF})&& \\ \nonumber +{\bf A}_{ij}^{-1}
\biggm( \frac{\delta \overline{v}_{xj}^S}{\delta u_i}
 -\frac{\delta \overline{v}_{xi}^{HF}}{\delta u_i} \biggm) \biggm] 
&& \\ \nonumber
+ \frac{u_i^*}{\rho}
\sum_{j=1}^{m-1} {\bf A}_{ij}^{-1}
(\overline{v}_{xj}^S-\overline{v}_{xj}^{HF}),
\eeqn
\be
\frac{\delta \overline{v}_{xj}^S}{\delta u_i}=(\frac{\delta e_x^{HF}}{\delta u_i}
\rho-e_x^{HF} u_i^*) \rho^{-2} \rho_j.
\ee
The functional derivative over the HF exchange energy density
\be
\frac{\delta e_x^{HF}}{\delta u_i}
=-\frac{1}{4} \sum_{j=1}^{occ}
\int d{\bf r}' \frac{u_i^*({\bf r}')
u_j({\bf r}) u_j^*({\bf r}')}{\vert {\bf r}-{\bf r}' \vert}.
\ee
For the inverse matrix ${\bf A}_{ij}^{-1}$ the following equality holds,
\be
\frac{\delta {\bf A}_{ij}^{-1}}{\delta u_i}=
\sum_{j=1}^{m-1} {\bf A}_{ij}^{-1} \frac{\delta {\bf A}_{ij}}{\delta u_i}
{\bf A}_{ij}^{-1},
\ee
where
\be
\frac{\delta {\bf A}_{ij}}{\delta u_i}=
-\frac{u_i^* \rho_j}{\rho^2} (1- \rho_i).
\ee
For simplicity, we give here the terms in conjuction with the remarkable KLI approximation \cite{KLI},
therefore the last term given in Eq. (15) is neglected.
However, this term is rather small \cite{KLI,Graborev} and can be accounted for
when high numerical accuracy is required.


\small


\begin{thebibliography}{999}

\bibitem{DFT}
P. Hohenberg, W. Kohn, Phys. Rev. {\bf 136}, B864. (1964),
W. Kohn, L. J. Sham, Phys. Rev. {\bf 140}, A1133. (1965),
W. Kohn, P. Vashishta, {\em General Density Functional Theory},
in.: {\em Theory of the Inhomogeneous Electron Gas}, (Plenum Press,
New York, 1983),
R. M. Dreizler, E. K. U. Gross, {\em Density Functional Theory}
 Springer, Berlin, 1990),
\'A. Nagy, {\em Density Functional Theory and Applications to
Atoms and Molecules}, Physics Reports, {\bf 298}, 1. (1998)

\bib{Perdew}
K. Burke, J. P. Perdew, M. Ernzerhof, Int. J. Quant. Chem., {\bf 61},
287. (1997)\\

\bibitem{KLI}
J. B. Krieger, Y. Li, G. J. Iafrate, Phys. Rev. {\bf A45}, 101. (1992),
see also in: {\em Density Functional Theory}, p. 191, Ed. by E.K.U. Gross
and R. M. Dreizler, NATO ASI Series, 1995\\ 

\bibitem{Sule}
O. Gritsenko, R. van Leeuwen, E. van Lenthe, E. J. Baerends,
Phys. Rev. {\bf A51}, 1944. (1995),
P. S\"ule, O. Gritsenko, \'A. Nagy, E. J. Baerends, J. Chem. Phys.,
{\bf 103}, 10385., (1995),
P. S\"ule, Chem. Phys. Lett., {\bf 259}, 69. (1996)\\

\bib{March}
N. H. March, {\em Electron Correlation in Molecules and Condensed
Phases}, Plenum Press, London, 1996

\bib{Sharp}
R. T. Sharp, G. K. Horton, Phys. Rev. {\bf 90}, 317. (1953).\\

\bib{TS}
J. D. Talman, W. F. Shadwick, Phys. Rev. {\bf A14}, 36. (1976).

\bib{Grabo}
T. Grabo, E. K. U. Gross, Int. J. Quant. Chem., {\bf 64}, 95. (1997)\\

\bib{CS}
R. Colle, D. Slavetti, Theoret. Chim. Acta, {\bf 37}, 329. (1975)

\bib{Gritsenko2}
O. V. Gritsenko, R. Van Leeuwen and E. J. Baerends., J. Chem. Phys., 
{\bf 104}, 8535. (1996)

\bib{GL}
A. G\"orling, M. Levy, Phys. Rev. {\bf B47}, 13105. (1993),
{\em ibid}, {\bf A50}, 196. (1994),
{\em ibid}, {\bf B47}, 13105. (1995),
{\em ibid}, {\bf A52}, 4493. (1995),
S. Ivanov, R. Lopez-Boada, A. G\"orling, M. Levy, J. Chem. Phys.,
{\bf 109}, 6280. (1998)

\bib{Casida}
M. E. Casida, Phys. Rev. {\bf A51}, 2005. (1995)

\bib{Sham}
L. J. Sham, M. Schl\"uter, Phys. Rev. Lett., {\bf 51}, 1888. (1983)

\bib{Graborev}
T. Grabo, T. Kreibich, S. Kurth, and {E.K.U. Gross},  in {\em The {S}trong
  {C}oulomb {C}orrelations and {E}lectronic {S}tructure {C}alculations:
  {B}eyond {L}ocal {D}ensity {A}pproximations}, edited by V. Anisimov (Gordon
  and Breach, Amsterdam, to appear).

\bib{Engel}
E. Engel, A. F. Bonetti, S. Keller, I. Andrejkovics, H. M\"uller,
R. M. Dreizler,
Phys. Rev. {\bf A58}, 964. (1998),
T. Kreibich, E. K. U. Gross, E. Engel, Phys. Rev. {\bf A57}, 138. (1998),
E. Engel, S. Keller, A. Facco Bonetti, H. M\"uller, R. M. Dreizler,
Phys. Rev. {\bf A52}, 2750. (1995)

\bib{Engel2}
E. Engel, S. H. Vosko, Phys. Rev. {\bf A38}, 3098. (1993)

\bib{Talman}
K. Aashamar, T. M. Luke, J. D. Talman, J. Phys. B {\bf 14}, 803. (1981)

\bib{Gonze}
 X. Gonze, D. C. Allan, M. P. Teter, Phys. Rev. Lett. {\bf 68}, 3603.
(1992),
X. Gonze, Phys. Rev. {\bf A52}, 1086. (1995), {\em ibid} {\bf A52},
1096. (1995) 

\bib{HolasMarch}
A. Holas, N. H. March, Phys. Rev. {\bf A56}, 3597. (1997),
A. Holas, N. H. March, Int. J. Quant. Chem. {\bf 61}, 263. (1997)

\bib{Sulepolym}
P. S\"ule, S. Kurth, V. E. Van Doren, Phys. Rev. {\bf B60}, (1999)


\bib{Nagy}
\'A. Nagy, Phys. Rev. {\bf A55}, 3465. (1997)

\bib{Perdew2}
V. Sahni, J. Gruenebaum, J. P. Perdew, Phys. Rev. {\bf B26}, 4371. (1982),
J. P. Perdew, M. R. Norman, Phys. Rev. {\bf B26}, 5445. (1982)

\bibitem{Slater}
 J. C. Slater, Phys. Rev. {\bf 81}, 385. {1951}

\bib{Shaginyan}
V. R. Shaginyan, Phys. Rev. {\bf A47}, 1507. (1993)

\bib{Stadele}
M. St\"adele, J. A. Majewski, P. Vogl, A. G\"orling, Phys. Rev. Lett.,
{\bf 79}, 2089. (1997)

\bib{Kotani}
T. Kotani, Phys. Rev. Lett. {\bf 74}, 2989. (1995)

\bib{Bylander}
D. M. Bylander, L. Kleinman, Phys. Rev. Lett. {\bf 74}, 3660. (1995),
{\bf 75}, 4334. (1995)

\bib{GL2}
A. G\"orling, M. Levy, Int. J. Quant. Chem. {\bf 29}, 93. (1995)

\bib{MP}
C. M$\o$ller, M. S. Plesset, Phys. Rev. {\bf 46}, 618. (1934)\\

\bib{Szabo}
A. Szab\'o, N. S. Ostlund, {\em Modern Quantum Chemistry} (McGraw-Hill,
New York, 1989), S. Wilson, {\em Electron Correlation in Molecules},
(Clarendron Press, Oxford, 1984)\\

\bib{Warken}
M. Warken, Chem. Phys. Lett., {\bf 237}, 256. (1995)

\bib{FBartha}
F. Bartha, {\em private communication}

\bib{PerdewLevy}
J. P. Perdew, M. Levy, Phys. Rev. Lett., {\bf 51}, 1884. (1983)

\bib{Chaudhuri}
R. K. Chaudhuri, J. P. Finley, K. F. Freed,
J. Chem. Phys., {\bf 106}, 4067. (1997), and references therein

\end{thebibliography}

\end{document}